\documentstyle[11pt,newpasp,twoside,epsf]{article}
\def\edcomment#1{\iffalse\marginpar{\raggedright\sl#1\/}\else\relax\fi}
\reversemarginpar

\begin{document}
\title{AMiBA: Array for Microwave Background Anisotropy}
\author{K. Y. Lo, T. H. Chiueh}
\affil{Academia Sinica Institute of Astronomy $\& $ Astrophysics (ASIAA) and Physics department, 
National Taiwan University (NTU)}
\author{R. N. Martin, Kin-Wang Ng}
\affil{ASIAA}
\author{H. Liang}
\affil{Physics department, University of Bristol}
\author{Ue-li Pen}
\affil{Canadian Institute of Theoretical Astrophysics (CITA) and ASIAA}
\author{Chung-Pei Ma}
\affil{Physics department, University of Pennsylvania and ASIAA}

\begin{abstract}
As part of a 4-year Cosmology and Particle Astrophysics (CosPA)
Research Excellence Initiative in Taiwan, AMiBA $-$ a 19-element
dual-channel 85-105 GHz interferometer array is being specifically
built to search for high redshift clusters of galaxies via the
Sunyaev-Zeldovich Effect (SZE).  In addition, AMiBA will have full
polarization capabilities, in order to probe the polarization
properties of the Cosmic Microwave Background.  AMiBA, to be sited
on Mauna Kea in Hawaii or in Chile, will reach a sensitivity of $\sim 1$\,mJy
or 7$\mu$K in 1 hour.  The project involves extensive international 
scientific and technical collaborations.  The construction of AMiBA 
is scheduled to starting operating in early 2004.
\end{abstract}

\section{Introduction}

The Academia Sinica Institute of Astronomy $\& $ Astrophysics and the
National Taiwan University Physics department in Taipei, Taiwan are jointly
developing experimental and theoretical Cosmology.  

The idea of a ground-based millimeter-wave interferometric array
(MINT) to study the primary anisotropy of the Cosmic Microwave
Background (CMB) was first suggested by Lyman Page of Princeton
University at a workshop on Cosmology held in Taiwan in December 1997.
Subsequently, the ASIAA focussed on designing an instrument
specifically for observing the Sunyaev-Zel'dovich Effect (SZE) to study
clusters of galaxies and to search for high redshift clusters to take
advantage of the distance-independent nature of the SZE.  The resulting
specifications consist of a high sensitivity 19-element 90 GHz
interferometer that can achieve one arc-minute resolution.

In response to the Research Excellence Initiative of the Ministry of
Education and the National Science Council in Taiwan, a proposal on
Cosmology and Particle Astrophysics (CosPA) was jointly submitted by
the NTU Physics department, the ASIAA, the National Central University
and the National Tsinghua University during the Spring of 1999, with
the goals of developing cosmological research and Optical/Infrared
Astronomy.  CosPA consists of five inter-related projects: (1)
construction and use of the Array for Microwave Background Anisotropy
(AMiBA); (2) theoretical work in Cosmology; (3) access to large
Optical/Infrared (OIR) facilities; (4) improving the infra-structure
at the Lulin observatory site in the Jade Mountains in Taiwan; (5) a
feasibility study of cold dark matter detection.

During December of the same year, the 4-year US$\$ $15M CosPA proposal
was funded in full.  In February 2000, a science and engineering
specification meeting on AMiBA was held in order to define an
instrument that will have unique scientific capabilities when
completed in late 2003.  The important conclusion was reached to achieve
full polarization capability for AMiBA in order to probe the
polarization properties of the CMB, in addition to observing the
secondary anisotropy of the CMB.

\section{Science Goals}

There are three principal science goals for AMiBA: (1) a survey for
high z clusters via the SZE; (2) in search of the missing baryons in
large scale structures via the SZE; and (3) the polarization of the
CMB.

\subsection{High z Cluster Survey}

The formation history of clusters depends on $\Omega_m$, the matter
density and $\Lambda$, the cosmological constant (e.g. Barbosa et
al. 1996), as well as $\sigma_8$, a measure of the initial
fluctuation amplitude (Fan \& Chiueh 2000).
What is needed observationally to define the history is a
survey of high z clusters over a sufficiently large area of the sky,
so that the cosmic variance does not affect the results
significantly. Because the SZE is distance independent, surveying for
the SZ decrement in the CMB is very well suited to search for clusters
at high z (Sunyaev \& Zel'dovich 1972; Birkinshaw 1998).

To optimize the sensitivity of the survey, AMiBA is designed for
maximal sensitivity, by maximizing the number of elements, adopting
dual channel for the receivers and to have a 20 GHz bandwidth.  The
choice of the 90 GHz range is to minimize the foreground and point
source confusion, and to minimize the scale of the array.

To detect clusters more massive than $2.5\times 10^{14}$\,M$_{\sun}$ at z
$\ga$ 0.7, the AMiBA will be more sensitive compared to X-ray
detection by satellites such as the XMM (fig. 1).  A comparison of
the sensitivity of AMiBA with other existing and planned instruments
that can be applied to such a survey is also shown in fig. 1. We plan to 
survey for 50 square degrees of sky at a speed of 1.2 square degree
~per month.

\begin{figure}
\plottwo{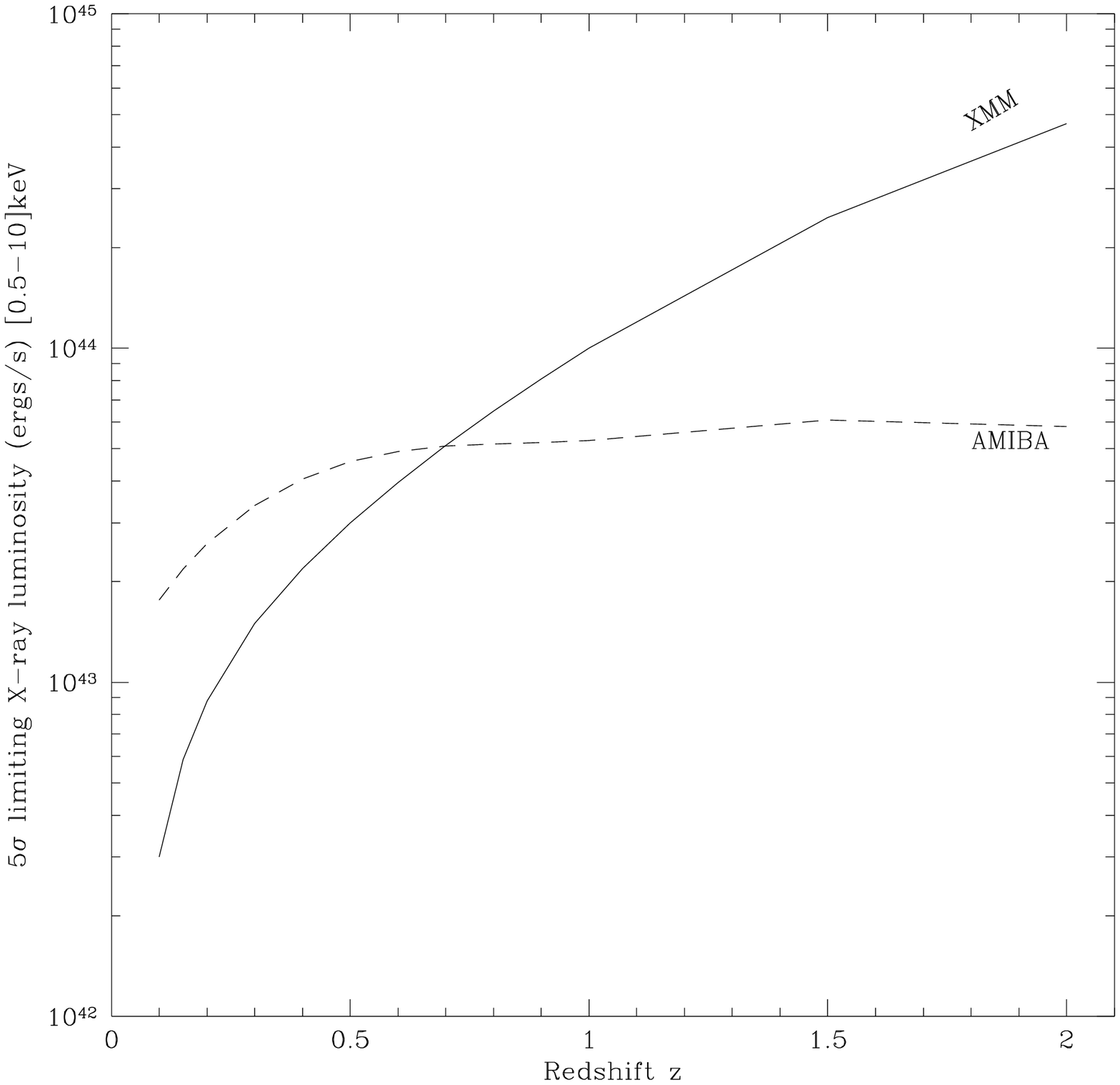}{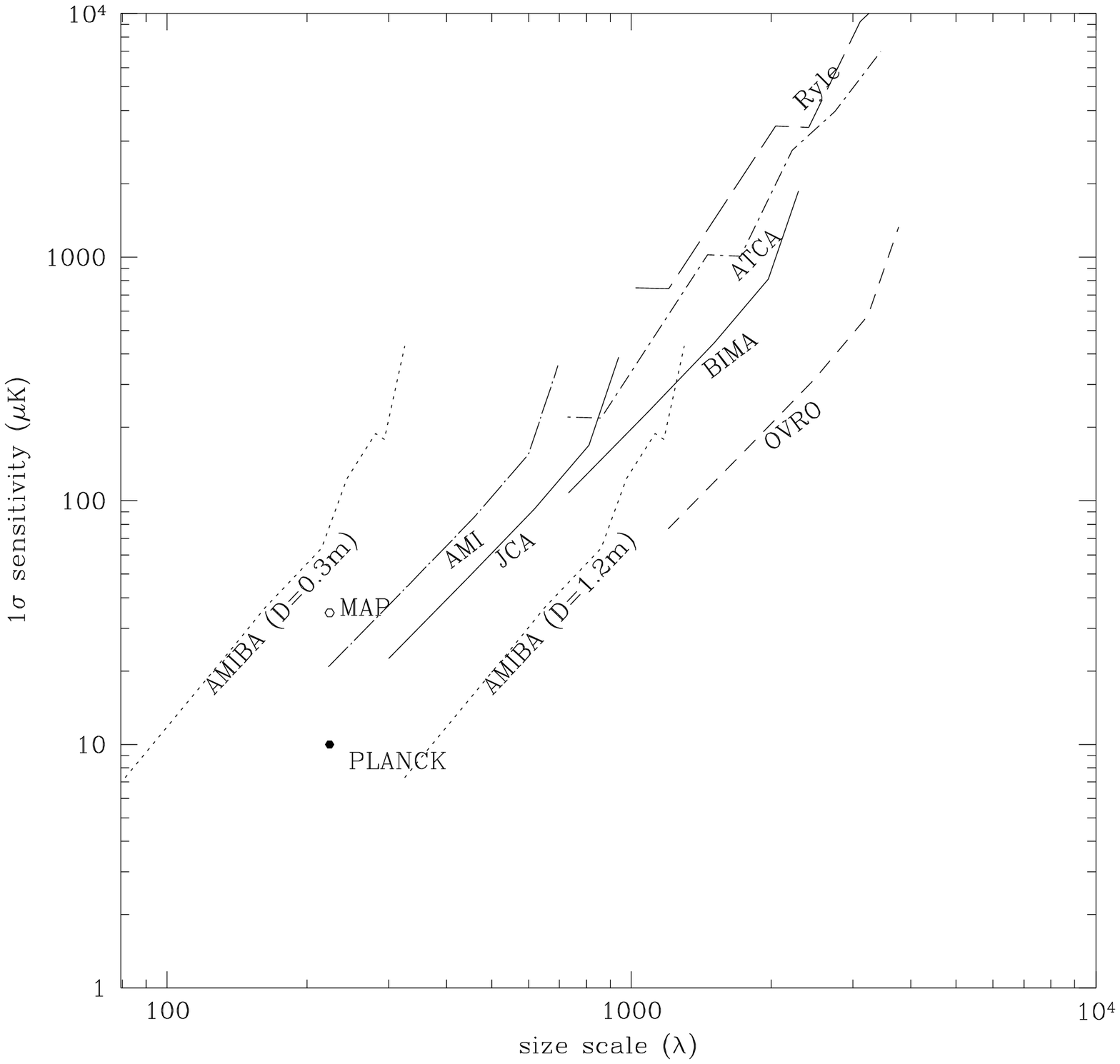}
\caption[]{LEFT:A comparison of the limiting sensitivity of AMIBA (dotted
curve) and XMM (Solid curve) for detecting clusters at various
redshifts.  AMIBA is more sensitive than XMM at detecting clusters
beyond a redshift of $z \ga 0.7$.  Limiting X-ray luminosity in
ergs~s$^{-1}$ for a 5$\sigma$ detection in 20 ksec for both telescopes
for a typical cluster of core radius $r_{c}=250$ kpc, shape parameter
$\beta=2/3$ and gas temperature $T_{e}=8$ keV, assuming an average
galactic neutral hydrogen column density of $N(H)=6\times10^{20}$
cm$^{-2}$ for the X-ray observation.  (XMM sensitivities provided
by Monique Arnaud) RIGHT: A comparison of brightness sensitivities (in 1hr) of
various ground based aperture synthesis telescopes for the SZE either
in operation or planning. The telescopes plotted is by no means a
complete list of telescopes capable of detecting the SZE. The
brightness sensitivity in $\mu$K over 1hr's observation is plotted
against the uv-spacing (or size scale) expressed in terms of the
number of wavelengths (multiplied by $2\pi$ gives $l$). A hexagonal
close packed configuration is assumed for AMIBA, and a scaled version
of a standard BIMA D-array configuration is assumed for the
non-platform arrays BIMA, OVRO, AMI and JCA. }
\end{figure}

\subsection{Super-clusters, Filaments - Missing Baryons ?}

In addition to the hot gas in highly collapsed objects such as rich
clusters, a weaker SZ effect should be produced when CMB photons
scatter off warm baryons in lower-density environments such as
filaments and inter-cluster regions in superclusters.  There is a
growing consensus that a significant fraction (from 1/3 to 1/2) of the
present-day baryons from big bang nucleosynthesis may be in the form
of warm to hot gas with $10^5 < T < 10^7$ K which has mostly eluded
detection thus far (cf. Fukugita et al 1998).  Both gravitational
and non-gravitational (such as supernova feedback) heating mechanisms
have been discussed for this gas component (Dave et al. 2000; Cen \&
Ostriker 1999; Pen 1999; Wu et al. 1999) The gravitational case is due
to shock heating as intergalactic gas flows along dark matter in
filaments and the large scale structure.  The dark matter in
these regions are at moderate overdensities with a mean of 10 to 30,
and the expected SZ distortion is of order 10 $\mu K$.  Detection of
signals from sensitive, non-targeted SZ surveys over large regions of
the sky may be feasible, but it will be challenging to separate out
the warm gas component from the primary anisotropy and the hot gas in
clusters.  We are currently carrying out more detailed studies to
assess this exciting possibility.

\subsection{Polarization of the CMB}

The CMB polarization contains a wealth of information about the early 
Universe.  It provides a sensitive test of the reionization history
as well as the presence of non-scalar metric perturbations, and improves
the accuracy in determining the cosmological parameters~(Zaldariaga et al 1997).
The degree of polarization is of order a few $\mu$K at $l\sim 1000$ 
(Bond and Efstathiou 1984). 

So far, the current upper limit on the CMB linear
polarization is $16~\mu$K~(Netterfield et al 1995). 
A handful of new experiments,
adopting low-noise receivers as well as long integration time per
pixel, are underway or being planned~(Staggs et al 2000).
The MAP mission, launched in 2001, will be sensitive to the temperature-polarization 
correlation. The balloon-borne Boomerang and
Maxima experiments are scheduled flights in 2001 to measure polarization
with an angular resolution of $l<800$ and pixel sensitivity
of a few $\mu$K.  The ESA space mission Planck will have sensitivity to CMB polarization, but 
the mission is not scheduled for launch until 2007.

An interferometer array is very attractive for CMB observations in 
that it directly measures the power spectrum.
In addition, many systematic problems that are inherent in single-dish
experiments, such as ground and near field atmospheric pickup, and
spurious polarization signal, can be reduced or avoided in interferometry
(cf. White et al 1999). Balloon-borne experiments are usually plaqued by 
pointing accuracy.

The AMiBA, with dual-channel receivers and 4 correlators, will be able
to measure all four Stokes parameters simultanesously, so that the
array will be much more sensitive to detecting CMB polarization than
existing arrays, such as the Very Small Array (VSA), Degree Angular Scale
Interferometer (DASI), Cosmic Background Imager (CBI).  The AMiBA, when used 
with the $0.3\;{\rm m}$ apertures, will be sensitive to 
CMB polarization over the range $700<l<2000$. The S/N ratio in polarization
in 24 hr is about 4 at $l\sim 700$, and about 2 at $l\sim 1150$.

\section{AMiBA Specifications}

\begin{table}

\begin{tabular}{|l|l|}  \hline
Frequency ($\nu$) & 85-105 GHz\\ \hline
Bandwidth ($\delta\nu$) & 20 GHz \\ \hline
Polarisations ($N_{p}$) & 2-linear (XX,YY) \\ \hline
Receiver type & HEMT, cooled to 15 K \\ \hline
System Temperature  & 70 K \\ \hline
Number of Antennas ($N$) & 19 \\ \hline
Size of antennas ($D$) & 2 sets; 0.3m,1.2 m \\ \hline
Number of Baselines & 171 \\ \hline
Primary beam & 11$^{'}$,44$^{'}$ FWHM \\ \hline
Synthesized beam (full range) & $1^{'}$ - $19^{'}$ \\ \hline
Fequency bands & 8 chunks over 20\,GHz \\ \hline
Flux Sensitivity & 1.3mJy, 20.5\,mJy in 1hr \\ \hline
Brightness sensitivity  & 7$\mu$K in 1hr \\ \hline
Mount & Hexapod Mount: 3 rotational axes used \\ \hline
Platform & CFRP structure - 3 fold symmetry \\ \hline
\end{tabular}
\end{table}

\section{Organization of AMiBA}

The AMiBA project is a collaboration principally between ASIAA/NTU and
the Australia Telescope National Facility (ATNF), with important
participation by scientists elsewhere. K. Y. Lo is the
PI, with Robert Martin as the project manager, T. H. Chiueh
as the project scientist, Paul Shaw (NTU/ASIAA) as the
project administrator, Michael Kesteven (ATNF) as the system
scientist.  The other science and engineering team members include Ron
Ekers, R. Sault, M. Sinclair, Ravi Subrahmanyan and W. Wilson from the
ATNF, M. T. Chen, Y. J. Hwang, Kin-wang Ng from the ASIAA,
T. D. Chiueh, T. Chu, and H. Wang from NTU, Haida Liang (Bristol),
Chung Pei Ma (Penn/ASIAA), Ue-li Pen (CITA/ASIAA), Jeff Peterson
(CMU), and John Payne (NRAO).

\section{Some AMiBA Technical Details}

The dual channel 85-105 GHz receievers will be based on the MIC InP
HEMT amplifiers supplied by the National Radio Astronomy Observatory,
with similar specifications built for the MAP project (Popieszalski
2000).  The local oscillator system will be based on photonic devices
with fiber-optic transmission lines, which will minimize component
counts and make the distribution more stable.  The 20 GHz bandwidth
poses considerable technical chanllenges that are being met by a
17-lag analog correlator.  There will be four correlators built to
provide full polarization capabilities for AMiBA.  The 19 apertures
will be mounted on three 2.5m platforms supported on hexapod mounts.

As the project is also funded to develop the research capabilities of the 
universities in Taiwan, there are parallel development projects on the 
InP and GaAs MMICs that are aimed at satisfying the requirements of the AMiBA.
However, these development efforts are not placed on the critical paths 
of the AMiBA construction.

\section{Schedule of events}

A preliminary design review meeting was held in July 2000 in Taipei,
where the decision was made to build a prototype by September 2001 to
test the basic concepts and specifications.  After the proving of
concepts, the full system will be built to be completed in late 2003 
and to start observations in early 2004.

To further review the science goals and to keep up with the latest
development in this rapidly evolving field, an international workshop
in Taiwan on AMiBA-related science goals is being planned for June
2001.


\begin{references}

\begin{quote}

\verb"Barbosa, D., Bartlett J. G., Blanchard, A., Oukbir J., 1996," \\
\verb"A&A, 314,13"\\
\verb"Birkinshaw, M., 1999, Phys. Rept., 310, 97 "\\
\verb"Bond, J. R., Efstathiou, G. 1984, ApJ, 285, L45"\\
\verb"Cen, R., Ostriker, J., 1999 ApJL, 519, L109 "\\
\verb"Dave,  et al 2000, astro-ph{0007217}"\\
\verb"Fan, Z, Chiueh, T. 2000, ApJ (in press), astro-ph/0011452"\\
\verb"Fukugita, M., Hogan, C. J., Peebles, P. J. E. 1998, ApJ, "\\ 
\verb"53, 518"\\
\verb"Netterfield, C. B. et al. 1995, ApJL, 474, L69"\\
\verb"Pen, U-L, 1999, ApJL, 510, 1L"\\
\verb"Popieszalski, M. et al. 2000, IEEE MTT-S Symp. Digest,"\\
\verb" in press"\\
\verb"Staggs, S. T., Gundersen, J. O., Church, S. E. astro-ph/9904062"\\
\verb"Sunyaev, R.A., Zel'dovich, Ya. B., 1972, Comm. Astrophys."\\
\verb"Sp. Phys., 4, 173"\\
\verb"White, M., Carlstrom,  J. E., Dragovan, M., Holzapfel, W. L."\\
\verb"1999, ApJ, 514, 12"\\
\verb"Wu, K. K. S., Fabian, A. C. Nulsen, P., 1999, MNRAS, in press "\\
\verb"Zaldarriaga, M., Spergel, D. N., Seljak, U. 1997, ApJ, 488, 1"\\

\end{quote}

\end{references}
\end{document}